\documentclass[aps,prl,twocolumn,groupedaddress,reprint,showpacs]{revtex4-2}
\usepackage{graphicx}
\usepackage{setspace}
\usepackage{amsmath}
\usepackage{amssymb}
\usepackage{color}
\usepackage{array}
\usepackage{subfigure}
\usepackage{hyperref}
\usepackage{float}
\usepackage{lipsum}

\usepackage[all]{xy}
\newcommand{\RN}[1]{%
  \textup{\uppercase\expandafter{\romannumeral#1}}%
}

\begin{document}

\title{Off-resonant modulated driving gate protocols for two-photon ground-Rydberg transition and finite Rydberg blockade strength}
\author{Yuan Sun}
\email[email: ]{yuansun@siom.ac.cn}
\affiliation{Key Laboratory of Quantum Optics and Center of Cold Atom Physics, Shanghai Institute of Optics and Fine Mechanics, Chinese Academy of Sciences, Shanghai 201800, China}

\begin{abstract}
Recently, the notion of two-qubit controlled phase gate via off-resonant modulated driving has been introduced into the neutral atom qubit platform, with respect to both single-photon and two-photon ground-Rydberg transitions. In order to reach a better performance practically, further developments are in need to overcome a few known limitations in previous discussions of this promising method. Here, we thoroughly analyze a variety of modulation styles for two-photon transitions, demonstrating the versatility of off-resonant modulated driving protocols. Furthermore, we show that it is possible to refine the designing process for improved performances for specific finite Rydberg blockade strength values. In particular, a reduced requirement on the blockade strength can be directly linked to an improvement of connectivity in qubit array of neutral atoms. These progress are closely related to the core feature that the atomic wave function acquires a geometric phase from the time evolution, which begins and finishes at the same quantum state. Under reasonable experimental conditions readily available nowadays, we anticipate that the fidelity of such protocols can reach as high as the essential requirement of NISQ even if the effects of technical errors and cold atoms' nonzero temperatures are considered.
\end{abstract}
\maketitle

%%%%

Rydberg atoms have attracted a lot of attentions not only in quantum information processing \cite{RevModPhys.82.2313}, but also in quantum sensing \cite{RevModPhys.89.035002}. The role of Rydberg atoms is of essential importance in the neutral atom qubit platform, as a consequence of the Rydberg blockade effect which can entangle two cold atoms at the distance of several to several tens of micrometers \cite{nphys1178, PhysRevA.92.022336}. The neutral atom qubit platform has seen many changes in recent years, thanks to several interesting technical progress, such as driving highly coherent ground-Rydberg transitions \cite{PhysRevLett.121.123603, Lukin2019arXiv}, building sizable neutral atom array \cite{Saffman2019arXiv, PhysRevLett.128.083202} and so on. Many efforts have also been devoted to quantum algorithms for this platform \cite{nphys41567}. While a lot of important problems are still awaiting for solutions, at this moment one of the most pressing tasks in this field is to improve the Controlled-PHASE gate fidelity to the level of Noisy Intermediate Scale Quantum (NISQ) technologies. Therefore, it raises major challenges on gate protocols to satisfy the highly-demanding requirement of practically reaching fidelity $\gtrsim 99.9\%$ without post-selection in experimental implementations with reasonably available apparatus. 

Recently, it has been discovered that the method of off-resonant modulated driving (ORMD) can be applied to construct a unique style of Controlled-PHASE gate \cite{PhysRevApplied.13.024059}, typically with symmetric driving on both the control and target qubit atoms. In particular, it can work with commonly available methods of coherently exciting an atom into the Rydberg levels, in terms of both the single-photon and two-photon ground-Rydberg transitions. Immediate experimental progress of two-qubit gates have demonstrated the feasibilities and advantages of ORMD protocols \cite{PhysRevA.105.042430}. Although some other adiabatic protocols can sometimes become unfriendly to experimental implementations due to their relatively long interaction times, a lot of interesting discussions have taken places about adiabaticity in the Rydberg blockade gates \cite{PhysRevA.101.062309, PhysRevA.103.012601, PhysRevApplied.17.024014, Shi_2022}, after the concepts of ORMD protocols were introduced.

So far, discussions about ORMD protocols for two-photon transitions are mostly limited to only modulating the driving optical field which couples the ground and intermediate levels. Moreover, the design of ORMD protocols used to focus on the idealized scenario where the Rydberg blockade strength can be generically regarded as infinitely large, which may not lead to an optimal performance at realistically finite value of Rydberg blockade strength, as we will discuss later. In this letter, we first discuss the broad possibilities of modulation styles in implementing ORMD with two-photon transitions, and analyze the properties of several representative choices. Then we move on to tackle the problem that the finite values of Rydberg blockade strength adversely affect the performance ORMD protocols in general. In particular, the ORMD protocols can be tailored to have significantly improved performance at reduced values of Rydberg blockade strength, after essential changes are made in the designing process. Furthermore, we will show a few detailed examples to demonstrate the differences caused by the methods introduced in this letter.

Without loss of generality, the discussions here are devoted to the two-qubit system composed of alkali atoms such as Rb or Cs, and the qubit register states $|0\rangle, |1\rangle$ are encoded in the typical way, namely the magnetic-insensitive Zeeman sub-states of the two ground hyperfine levels. The relevant states of the two-photon ground-Rydberg transition form a three-level ladder system in terms of $|1\rangle\leftrightarrow|e\rangle\leftrightarrow|r\rangle$. The lasers interact with the two qubit atoms symmetrically such that each atom effectively receives the same pulse. The four basis states $|00\rangle, |10\rangle, |01\rangle, |11\rangle$ correspond to different types of time evolutions as initial states. After the rotating wave approximation, the Hamiltonian $H_{10}/\hbar$ for $|10\rangle$ is:
\begin{equation}
\label{eq:H10}
\frac{\Omega_p}{2}|10\rangle\langle e0| + \frac{\Omega_S}{2}|e0\rangle\langle r0| + \text{H.c.} 
+ \Delta|e0\rangle\langle e0| + \delta|r0\rangle\langle r0|
\end{equation}
with time-dependent Rabi frequencies $\Omega_p(t), \Omega_S(t)$ and fixed values of one-photon detuning $\Delta$ and two-photon detuning $\delta$. The situation of $|10\rangle$ is similar. Generally, to cut down population in the lossy state $|e\rangle$ the parameters are chosen as $\Delta \gg \Omega_p, \Omega_S$ and we neglect decays of $|e\rangle$.

On the other hand, the Hamiltonian $H_{11}$ for $|11\rangle$ consists of two parts: the atom-light interaction part of $H_I$ and the F\"orster resonance part of $H_F = B|rr\rangle \langle pp'| + \text{H.c.} + \delta_p|pp'\rangle\langle pp'|$ describing the Rydberg dipole-dipole interaction. Let $|\tilde{e}\rangle = (|e1\rangle+|1e\rangle)/\sqrt{2}$, $|\tilde{r}\rangle = (|r1\rangle+|1r\rangle)/\sqrt{2}$ and $|\tilde{R}\rangle = (|re\rangle+|er\rangle)/\sqrt{2}$, after the rotating wave approximation, Morris-Shore transform and neglecting the influence of $|ee\rangle$, $H_{11}/\hbar$ can be expressed as follows:
\begin{eqnarray}
&\frac{\sqrt{2}\Omega_p}{2}|11 \rangle \langle \tilde{e}| + \frac{\Omega_S}{2} |\tilde{e}\rangle\langle \tilde{r}| + \frac{\Omega_p}{2} |\tilde{r}\rangle\langle \tilde{R}| + \frac{\sqrt{2}\Omega_S}{2} |\tilde{R}\rangle \langle rr| + \text{H.c.} \nonumber\\
& + \Delta |\tilde{e}\rangle\langle \tilde{e}| + \delta |\tilde{r}\rangle\langle \tilde{r}|+ (\Delta+\delta) |\tilde{R}\rangle\langle \tilde{R}| + 2\delta|rr\rangle\langle rr| \nonumber\\
& + B|rr\rangle \langle pp'| + B|pp'\rangle \langle rr| + (2\delta + \delta_p)|pp'\rangle\langle pp'|.
\label{eq:H11}
\end{eqnarray}
Herein, $H_I$ corresponds to the linkage structure of $|11\rangle \leftrightarrow |\tilde{e}\rangle \leftrightarrow |\tilde{r}\rangle \leftrightarrow |\tilde{R}\rangle \leftrightarrow |rr\rangle$ while $H_F$ corresponds to $|rr\rangle \leftrightarrow |pp'\rangle$. The dipole-dipole coupling is set as a real number $B$, which is effectively also the Rydberg blockade strength, and the F\"{o}rster energy penalty term $\delta_p$ for $|pp'\rangle$ usually is relatively small. $|00\rangle$ stays unchanged as it does not participate the atom-laser interaction, and therefore the overall Hamiltonian for the system is effectively $H_\text{tot}= H_{01} + H_{10} + H_{11}$.

On a finite time interval, the Bernstein polynomials form a complete basis for $L^2$ functions, and therefore for practical convenience and theoretical reasonableness, a smooth pulse can be expressed by their linear superpositions. The main purpose is to represent continuous pulse functions by discrete numbers, which become more friendly to work with in the numerical optimization and verification process. The choice of basis is not unique, for example, Fourier series or wavelets can become alternative choices, which seem suitable to handle the special requirements on the frequency domain. Experimentally, a continuous pulse starts and ends at zero strength. Moreover, the terms of expansion cannot go to infinity, so that truncation is deemed as necessary. Therefore, the waveform of modulations reduces to the following format:
\begin{equation}
f_{N, T_g} (t) = \sum_{\nu=1}^{N-1} \alpha_\nu b_{\nu, N}(t/T_g),
\end{equation}
where $b_{\nu, n}$ represents the $\nu$th Bernstein basis polynomials of degree $n$, $\alpha_\nu$'s are coefficients of expansion and $T_g$ is the time reference.

\begin{figure}[t]
\centering
\includegraphics[width=0.49\textwidth]{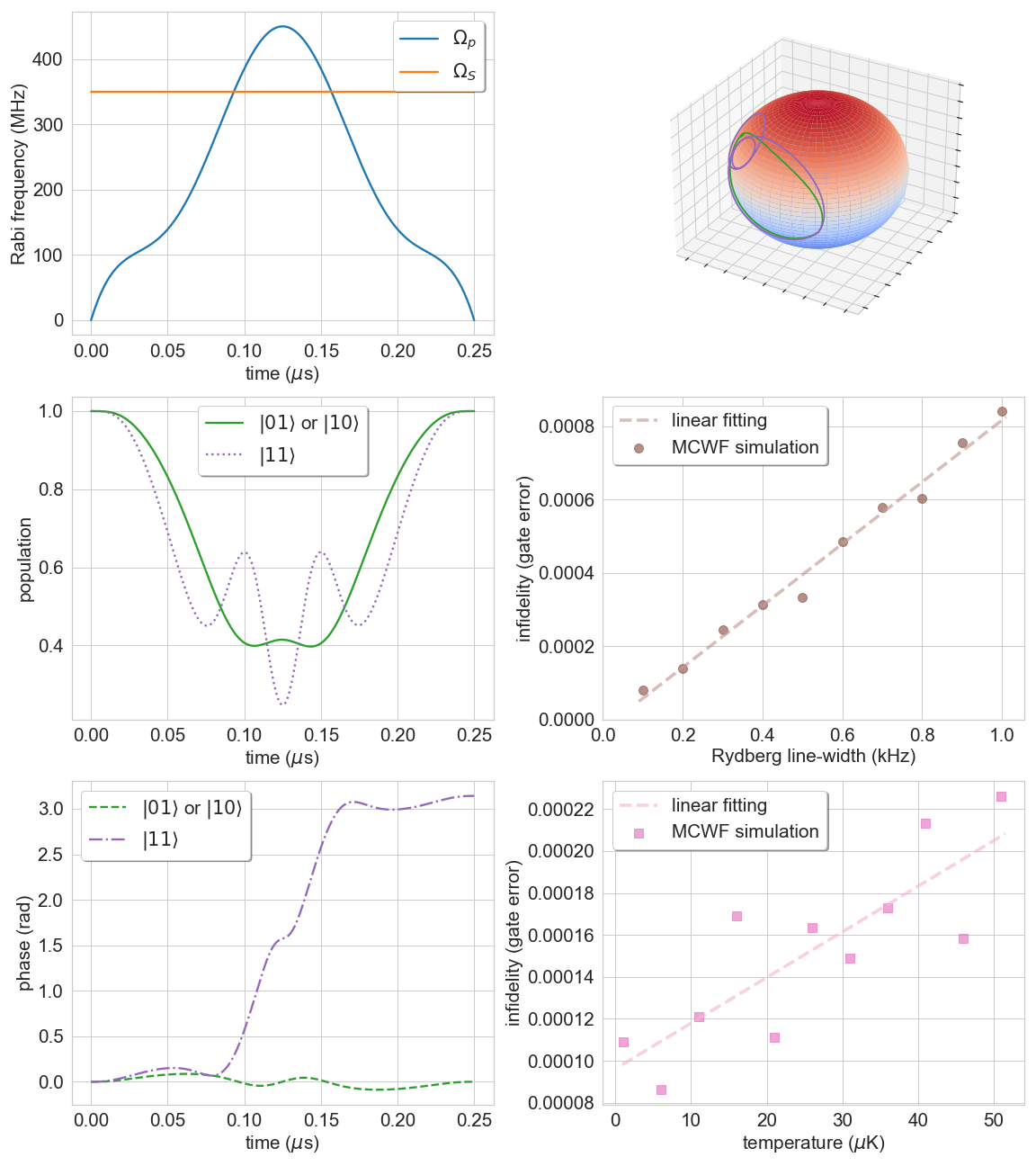}
\caption{(Color online) Controlled-Z gate via only modulating $\Omega_p$. (Left top) Waveforms of Rabi frequencies. (Left center) Time evolution of populations. (Left bottom) Time evolution of phases. (Right top) Time evolution of wave functions as trajectories on the Bloch sphere based on the ground level and singly-excited Rydberg level; green line for $|01\rangle, |10\rangle$ and purple line for $|11\rangle$. (Right center) Calculation of fidelities vs. line-widths of Rydberg states by MCWF. (Right bottom) Calculation of fidelities vs. temperatures of qubit atoms by MCWF with Rydberg line-width at $0.1$ kHz. The gate time here is 0.25 $\mu$s. $B/2\pi=500$ MHz and $\delta_p=0$ for $H_F$. Each point of MCWF numerical calculation is averaged from $1.5\times10^5$ trajectories.}
\label{fig1:type_A_waveform}
\end{figure}

In the limit of idealized Rydberg blockade effect with $B\to\infty$, $H_{11}$ decouples from $|rr\rangle, |pp'\rangle$. Then the design of laser waveforms need satisfy the requirement that the populations return to the initial qubit register states respectively and the phase accumulations satisfy Controlled-PHASE gate condition after the atom-laser interaction prescribed by $H_\text{tot}$, as we have accomplished previously for single-photon transition \cite{PhysRevApplied.13.024059}. While an analytical solution is often not accessible, technically we may try to get answers from numerical optimization processes. It is possible to as many solutions as without paying actual attention to the details of underlying physics. However, at this moment, we are particularly interested in several representative modulation configurations, which have close ties with experimental efforts. 

ORMD gate protocols for two-photon ground-Rydberg transitions can be approximately divided into four types for the purpose of categorization. \textit{Type A}: only modulating $\Omega_p$; \textit{Type B}: only modulating $\Omega_S$; \textit{Type C}: both $\Omega_p$ and $\Omega_S$ are modulated, but asynchronously; \textit{Type D}: both $\Omega_p$ and $\Omega_S$ are modulated synchronously with the same waveform. While each type has its individual special features, they share the common fundamental property that the atomic wave function acquires a geometric phase from the time evolution.

\begin{figure}[t]
\centering
\includegraphics[width=0.49\textwidth]{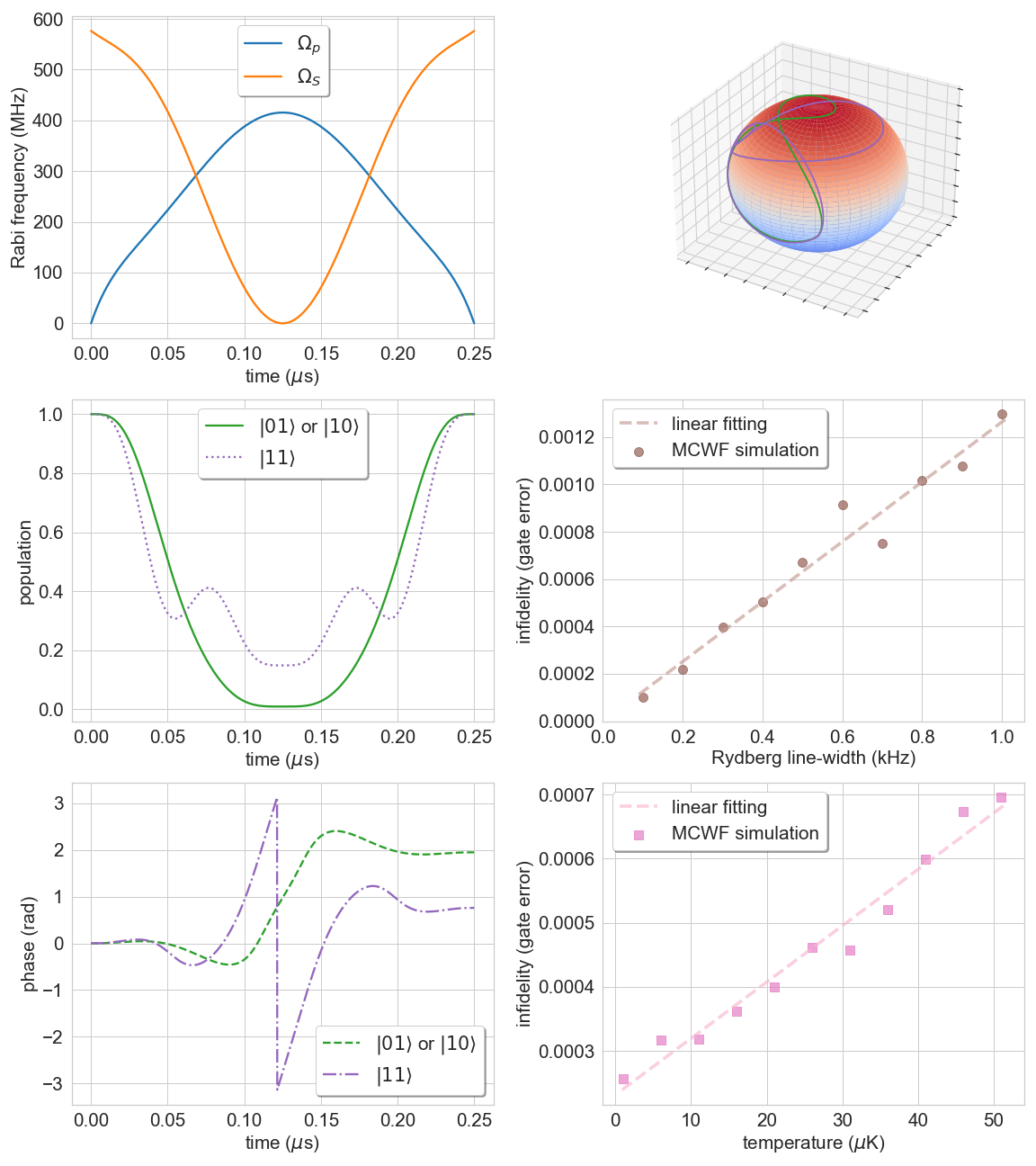}
\caption{(Color online) Controlled-PHASE gate via modulating both $\Omega_p, \Omega_S$ asynchronously. Arrangement of subplots are similar to that of Fig. \ref{fig1:type_A_waveform}. The gate time is 0.25 $\mu$s. $B/2\pi=500$ MHz and $\delta_p=0$ for $H_F$. Each point of MCWF numerical calculation is averaged from $1.5\times10^5$ trajectories.}
\label{fig2:type_C_waveform}
\end{figure}

An example of \textit{Type A} waveforms is shown in Fig. \ref{fig1:type_A_waveform}, and in particular it directly forms a Controlled-Z gate which saves the trouble of local phase corrections \cite{PhysRevA.92.022336, PhysRevA.105.042430}. Here $\Omega_p/2\pi$ is symmetrically modulated as $\sum_{\nu=1}^{5} \alpha_\nu \big(b_{\nu, 10}(t/T_g)+b_{10-\nu, 10}(t/T_g)\big)$ with $\alpha_1 = 193.65$ MHz, $\alpha_2 = 85.17$ MHz, $\alpha_3 = 0$, $\alpha_4 = 291.53$ MHz, $\alpha_5 = 649.10$ MHz and $T_g=0.25$ $\mu$s, and the other parameters include: $\Omega_S/2\pi = 350$ MHz, $\Delta/2\pi=5$ GHz and $\delta/2\pi=-0.5686$ MHz. The effects of finite lifetime of Rydberg levels and nonzero temperature of qubit atoms are also considered in the numerical simulation by the method of Monte-Carlo Wave Function (MCWF). The fidelity is evaluated by the method established in Refs. \cite{JAMIOLKOWSKI1972275, CHOI1975285, PhysRevA.71.062310, PEDERSEN200747}. As the results indicate, the lifetime of Rydberg levels is a major limiting factor for the ultimate fidelity of ORMD protocols apart from experimental imperfections. Meanwhile, it shows a reasonably robust performance against $\mu$K-scale temperatures, which benefits from the counter-propagating geometry between $\Omega_p, \Omega_S$ and the protocol's inherent property of avoiding actual photon absorption and reemission in the end. For \textit{Type B} waveforms, although one may as well pursue a Controlled-Z gate, but as a consequence of the ac Stark shift induced by non-zero value of $\Omega_p$ on the ground levels, it suffices to only consider Controlled-PHASE gate. Practically, $\Omega_p$ can be turned on and off with a fixed smooth ramping process, such that the ensued local phase correction can stay at a constant value. Further details are provided in the supplemental material.

An example of symmetric \textit{Type C} waveforms is shown in Fig. \ref{fig2:type_C_waveform}, building on $h(t/T_g) = \sum_{\nu=1}^{4} \alpha_\nu \big(b_{\nu, 8}(t/T_g)+b_{8-\nu, 8}(t/T_g)\big)$ with $T_g=0.25$ $\mu$s. More specifically, $\Omega_p/2\pi$ is set as $h(t/T_g)$ with $\alpha_1 = 277.99$ MHz, $\alpha_2 = 36.64$ MHz, $\alpha_3 = 649.60$ MHz, $\alpha_4 = 193.58$ MHz; and $\Omega_S/2\pi$ is set as $h(1/2) - h(t/T_g)$ with $\alpha_1 = 68.95$ MHz, $\alpha_2 = 51.10$ MHz, $\alpha_3 = 324.19$ MHz, $\alpha_4 = 766.36$ MHz. The other parameters include: $\Delta/2\pi=5$ GHz and $\delta/2\pi=4.462$ MHz. Its time evolution has obvious similarities with that of adiabatic rapid passage (ARP) between the ground and singly-excited Rydberg levels, as the intensities of $\Omega_p, \Omega_S$ vary asynchronously. It is anticipated that some of carefully designed \textit{Type C} protocols have enhanced robustness against experimental imperfections in waveforms.

\begin{figure}[b]
\centering
\includegraphics[width=0.49\textwidth]{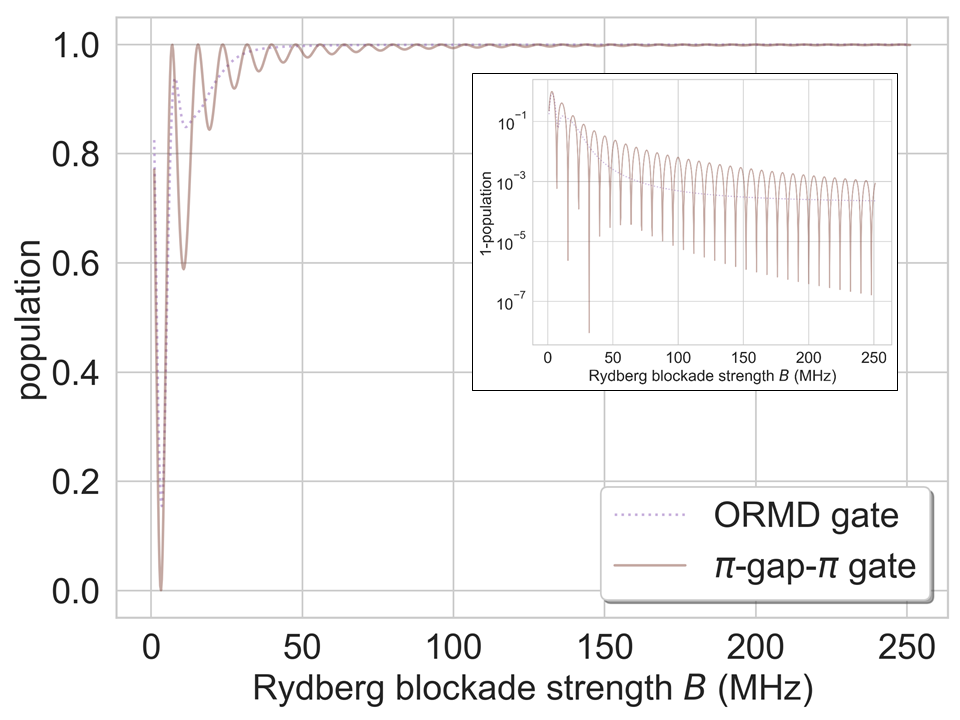}
\caption{(Color online) Performances of Controlled-Z gate protocols. The inset shows the same contents under log scale. The gate time is set as 0.25 $\mu$s.}
\label{fig3:scan_blockade_strength}
\end{figure}

\begin{figure}[b]
\centering
\includegraphics[width=0.49\textwidth]{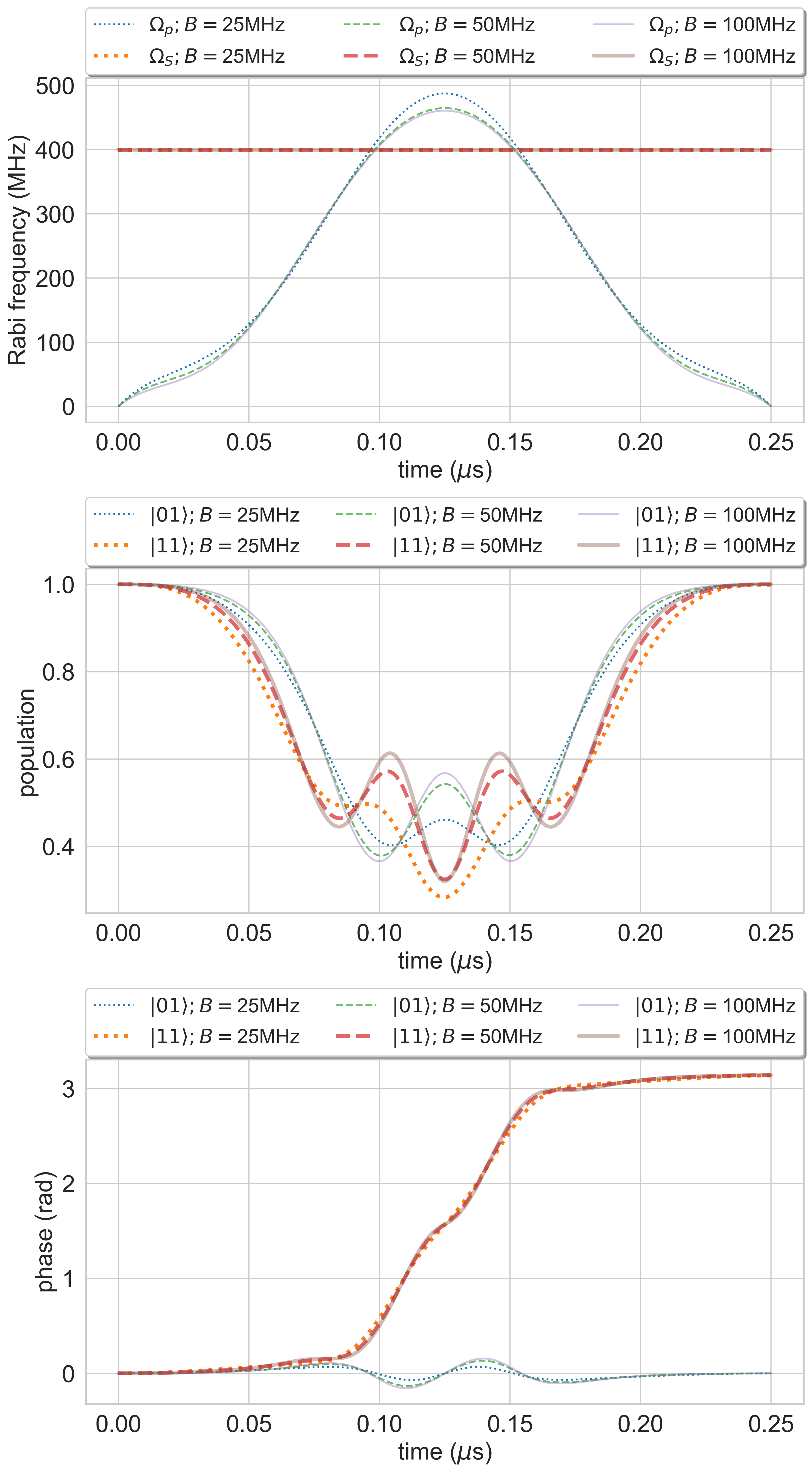}
\caption{(Color online) ORMD protocols for two-photon transitions designed with intention for Rydberg blockade strength. They are designed strictly according to Eq. \eqref{eq:H10} and Eq. \eqref{eq:H11}.}
\label{fig4:waveforms_ORMDwithIRBS}
\end{figure}

So far, the designing processes of ORMD protocols have always used the assumption of ideal Rydberg blockade effect. However, experiments practically encounter finite values of blockade strength $B$ instead of $B\to \infty$. Therefore, it makes sense to examine the performance of these protocols with respect to realistic values of $B$, and such a result is shown in Fig. \ref{fig3:scan_blockade_strength}. In particular, the time evolution of initial state $|11\rangle$ involving the Rydberg blockade effect is numerically calculated and the population in $|11\rangle$ after interaction is plotted. Fig. \ref{fig3:scan_blockade_strength} also includes the classic Controlled-Z protocol as a comparison with $\pi$-gap-$\pi$ pulse sequence on the control atom and $2\pi$ pulse on the target atom \cite{RevModPhys.82.2313}. The better performances of ORMD protocols indicate the advantageous mechanism of avoiding shelved population in the Rydberg levels and adiabatically tracking the dark state in the presence of Rydberg dipole-dipole interaction \cite{PhysRevApplied.13.024059}.  

Nevertheless, as revealed by Fig. \ref{fig3:scan_blockade_strength}, it still requires a relatively large value of $B$ to reach a theoretically high quality. Therefore this difficulty triggers the motivation to upgrade ORMD protocols to accommodate less than ideal Rydberg block strengths. It turns out, such a possibility is attainable in the sense that an ORMD protocol can be designed for specific values of $B, \delta_p$. The reasons can mostly be attributed to that a finite value of $B$ instead of $B\to\infty$ causes an additional phase shift to the first order. In other words, the Rydberg dipole-dipole interaction causes a consequence similar to dressing the two states $|rr\rangle$ and $|pp'\rangle$ with coupling strength $B$ and detuning $\delta_p$. From the viewpoint of dressed states, the transition to the doubly excited Rydberg level effectively experience an extra detuning of $\sqrt{B^2 + \delta_p^2}/2$, which results in an extra phase shift for the ORMD process. Although the detailed modeling of dipole-dipole interaction may vary, but the underlying physics does not fundamentally deviate from this observation. Therefore, if the values of $B, \delta_p$ of $H_F$ is taken as a known priori, specific waveforms can be accordingly obtained to retain high fidelities following the previously stated principles of designing ORMD protocols. An set of examples is shown in Fig. \ref{fig4:waveforms_ORMDwithIRBS} with $\delta_p=0$ for all of them. 

\begin{figure}[h]
\centering
\includegraphics[width=0.49\textwidth]{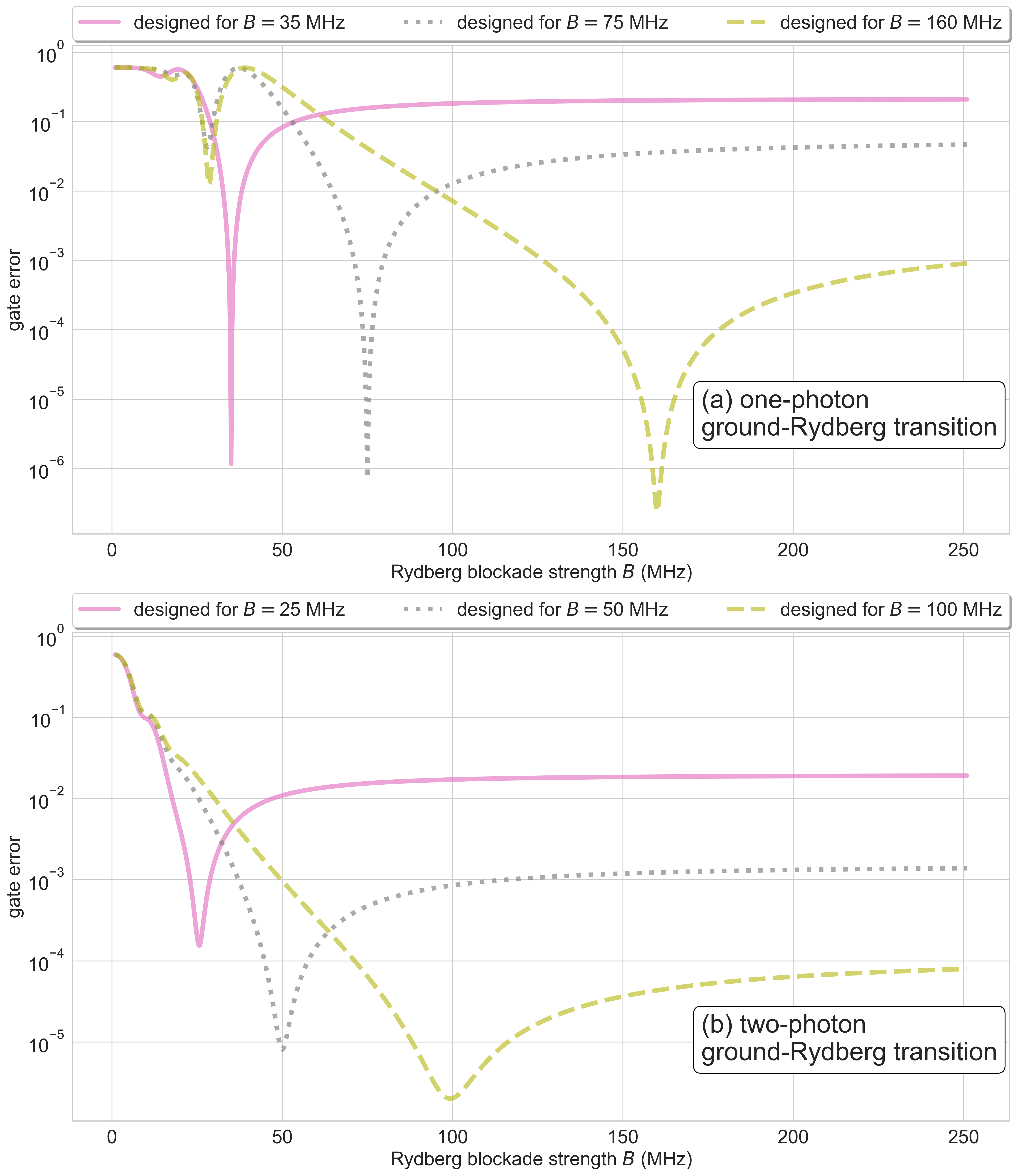}
\caption{(Color online) Performances of ORMD protocols with intention for Rydberg blockade strength. (a) Controlled-PHASE gate via one-photon transition. (b) Controlled-Z gate via two-photon transition.}
\label{fig5:properties_ORMDwithIRBS}
\end{figure}

While ORMD protocols can be designed specifically for $B, \delta_p$ now, the glaring question of performance against changes in $B$ still persists, because $B$ depends on the distance between atoms in typical Rydberg dipole-dipole interactions. Fortunately, ORMD protocols often have a certain degree of robustness in a neighborhood of the prescribed point, as demonstrated in Fig. \ref{fig5:properties_ORMDwithIRBS}, thanks to the inherent properties embedded in the special style of time evolution via ORMD. Moreover, it suggests that a direct implementation of Controlled-Z gate seems more preferable to Controlled-PHASE gate, when considering robustness against non-ideal Rydberg blockade and experimental imperfections. This subtle difference links with that the finite Rydberg blockade strength affects necessary local phase correction values for Controlled-PHASE gate. Overall, this upgrade can become especially helpful to enhance connectivity of cold atom qubit array. Namely, for a pair of relatively distant qubit atoms, one can choose a specific waveform corresponding to the dipole-dipole interaction strength to maintain a reasonable fidelity.

It has become a common practice to arrange the two lasers in a counter-propagating geometry for two-photon ground-Rydberg transitions, in order to reduce the Doppler-induced dephasing effects. Ideally, if the frequency differences of the two lasers are within a few percent, the cancellation can yield satisfactory behavior. Nevertheless, such a requirement can not be routinely fulfilled by the current mainstream experimental designs for alkali atoms. A more realistic strategy of dual-pulse technique towards an almost Doppler-free gate process can help with this issue \cite{PhysRevApplied.13.024059}.

In conclusion, we have reported the progress of designing two-qubit gate protocols for two-photon ground-Rydberg transition and finite Rydberg blockade strength via the method of ORMD. The extra degrees of freedom brought by two-photon transition allow for a rich variety of ORMD waveforms that can become helpful for the next stage of experimental efforts. Quite contrary to some of the previous speculations, ORMD protocols can still maintain a reasonably robust performance at a reduced value of Rydberg blockade strength, with essential modifications that do not increase the complexity of modulated waveforms. The derivations here assume symmetric driving and the same Rydberg state $|r\rangle$ on two qubit atoms throughout the discussions, but they apply as well to asymmetric drivings after appropriate modifications. We expect our work will also be helpful in precision measurement with Rydberg atoms, especially when ground-Rydberg Ramsey type pulse sequence is involved \cite{PhysRevLett.122.053601}. 

\begin{acknowledgements}
This work is supported by the National Natural Science Foundation of China (Grant No. 92165107 and No. 12074391), the National Key R\&D Program of China (Grant No. 2016YFA0301504), the Chinese Academy of Sciences and the China Manned Space Engineering Office. The author thanks Peng Xu and Xiaodong He for many discussions.
\end{acknowledgements}

\bibliographystyle{apsrev4-2}
\bibliography{rudder_ref}

\end{document}